\documentclass[acmlarge,nonacm]{acmart}
\makeatletter
\newcommand{\confshort}{\acmConference@shortname}
\newcommand{\conffull}{\acmConference@name}
\newcommand{\confdate}{\acmConference@date}
\newcommand{\confloc}{\acmConference@venue}
\AtBeginDocument{
  \fancypagestyle{firstpagestyle}{
    \fancyhead{}%
    \fancyfoot[C]{}%
  }
  \fancyhf{}
  \fancyhead[LO]{\@headfootfont\shorttitle}%
  \fancyhead[RE]{\@headfootfont\@shortauthors}%
  \fancyhead[LE]{\@headfootfont\footnotesize \confshort, \confdate, \confloc}%
  \fancyhead[RO]{\@headfootfont\footnotesize \confshort, \confdate, \confloc}%
  \fancyfoot[C]{}%
}
\makeatother
\acmBooktitle{\conffull\@ (\confshort), \confdate, \confloc}

\usepackage{graphicx} % Required for inserting images
\usepackage{subcaption}
\usepackage{enumitem}

\usepackage{booktabs}
\usepackage{tabularx}
\usepackage{ragged2e}

\usepackage{multirow}    % for \multirow
\AtBeginDocument{%
  }

\copyrightyear{2026}
\acmYear{2026}
\setcopyright{cc}
\setcctype{by}
\acmConference[FAccT '26]{The 2026 ACM Conference on Fairness, Accountability, and Transparency}{June 25--28, 2026}{Montreal, QC, Canada}
\acmBooktitle{The 2026 ACM Conference on Fairness, Accountability, and Transparency (FAccT '26), June 25--28, 2026, Montreal, QC, Canada}
\acmDOI{10.1145/3805689.3806436}
\acmISBN{979-8-4007-2596-8/2026/06}

\definecolor{bram}{rgb}{1, .75, 0}
\definecolor{lucas}{rgb}{0, 0, 1}
\definecolor{konrad}{rgb}{1, .44, .37}
\usepackage{tcolorbox}\usepackage{enumitem}

\usepackage{xcolor}

\title[Is your AI Model Accurate Enough?]{Is your AI Model \textit{Accurate} Enough? The Difficult Choices Behind Rigorous AI Development and the EU AI Act}
\author{Lucas G. Uberti-Bona Marin}
\email{lucas.uberti-bonamarin@maastrichtuniversity.nl}
\orcid{0000-0001-6518-9535}
\affiliation{%
  \institution{Law \& Tech Lab, Maastricht University}
  \city{Maastricht}
  \country{The Netherlands}
}

\author{Bram Rijsbosch}
\email{bram.rijsbosch@maastrichtuniversity.nl}
\orcid{0009-0000-0803-211X}
\affiliation{%
  \institution{Law \& Tech Lab, Maastricht University}
  \city{Maastricht}
  \country{The Netherlands}
}

\author{Kristof Meding}
\email{kristof.meding@uni-tuebingen.de}
\orcid{0000-0001-5073-2347}
\affiliation{%
  \institution{University of Tübingen - Computational Law Lab}
  \city{TÜBingen}
  \country{Germany}
}

\author{Gerasimos Spanakis}
\email{jerry.spanakis@maastrichtuniversity.nl}
\orcid{0000-0002-0799-0241}
\affiliation{%
  \institution{Law \& Tech Lab, Maastricht University}
  \city{Maastricht}
  \country{The Netherlands}
}

\author{Gijs van Dijck}
\email{gijs.vandijck@maastrichtuniversity.nl}
\orcid{0000-0003-4102-4415}
\affiliation{%
  \institution{Law \& Tech Lab, Maastricht University}
  \city{Maastricht}
  \country{The Netherlands}
}

\author{Konrad Kollnig}
\email{konrad.kollnig@maastrichtuniversity.nl}
\orcid{0000-0002-7412-8731}
\affiliation{%
  \institution{Law \& Tech Lab, Maastricht University}
  \city{Maastricht}
  \country{The Netherlands}
}


\date{January 2026}

\begin{document}
\begin{abstract}
Technical and legal debates frequently suggest that ``accuracy'' is an objective, measurable, and purely technical property. We challenge this view, showing that evaluating AI performance fundamentally depends on context-dependent normative decisions. These \textit{techno-normative} choices are crucial for rigorous AI deployment, as they determine which errors are prioritised, how risks are distributed, and how trade-offs between competing objectives are resolved.

This paper provides a legal-technical analysis of the choices that shape how accuracy is defined, measured, and assessed, using the 2024 European Union AI Act~--~which mandates an ``appropriate level of accuracy'' for high-risk systems~--~as a primary case study. We identify and analyse four choices central to any robust performance evaluation: (1) selecting metrics, (2) balancing multiple metrics, (3) measuring metrics against representative data, and (4) determining acceptance thresholds.
% Perhaps stress in the abstract that the techno-normative choices on a very down-to-earth level (not some high-level stuff, which you often see) is the novelty of this paper
For each choice, we study its relationship to the AI Act's accuracy requirement and associated documentation obligations, show how its technical implementation embeds implicit or explicit assumptions about acceptable risks, errors, and trade-offs, and discuss the implications for the practical implementation of the AI Act by examples and related technical standards. 

By making the techno-normative dimensions of accuracy explicit, this paper contributes to broader interdisciplinary debates on AI governance and regulation, and offers specific guidance for regulators, auditors, and developers tasked with translating (legal) safety requirements into technical practice.
\end{abstract}

\newcommand\blfootnote[1]{%
  \begingroup
  \renewcommand\thefootnote{}\footnote{#1}%
  \addtocounter{footnote}{-1}%
}

\maketitle
\section{Introduction}
%\begin{figure}[h]
%    \centering
%    \includegraphics[width=0.65\textwidth]{figures/Melanoma.png}
%    \caption{Melanoma detection tool, apparently with 99.8\% accuracy. Source: Associated Press~\cite{AI_Skin_Cancer_App_2025}}
%    \Description{Screenshot of an Associated Press Videos news article reporting on an AI skin cancer detection app claiming 99.8 percent accuracy rate in ruling out cancer }
%    \label{fig:Melanoma}
%\end{figure}
When is an AI model accurate enough to be deployed? 
Consider an AI-based skin cancer detection app that claims to detect melanoma with ``99.8\% accuracy''~\cite{AI_Skin_Cancer_App_2025}. %, as illustrated in Figure \ref{fig:Melanoma}.
At first glance, such a system might appear highly promising. Yet, because melanoma is extremely rare in the general population, a system that always predicts the absence of melanoma~--~when looking at the whole population~--~could achieve a similarly high accuracy score, while failing to be any useful. High accuracy, thus, on its own, does not necessarily imply meaningful or safe performance. 
This example illustrates some of the challenges in AI model performance evaluations (also referred to as accuracy evaluations, or model testing). The objective of performance evaluations is to generate quantitative evidence -- most commonly in the form of performance metric scores, such as accuracy -- about how well a model performs with respect to its intended purpose.

From a legal perspective, this stage thus plays a crucial role: it is where final decisions are made about whether a model’s performance is adequate for deployment, and where risks associated with the use of a model should be assessed and documented.
Recognising this importance, the 2024 EU AI Act introduced a legal requirement that directly targets the performance evaluation stage of AI model development, requiring that  
\textit{high-risk AI systems} (which an AI-based melanoma detector would likely qualify as) ``achieve an appropriate level of accuracy'' before they may be placed on the EU market \cite{EU_2024_AI_Act}, hereby aiming to ensure that only safe and trustworthy AI systems are used in the EU.

Performance evaluations are neither a trivial nor a purely technical exercise. They require making a series of interdependent design choices -- such as selecting and balancing performance metrics, defining test procedures, and setting performance thresholds -- that are often highly context-dependent and normatively-laden: They encode explicit or implicit assumptions about what errors are acceptable, which risks are prioritised, and how trade-offs are taken into account by a model.
In the case of an AI-based melanoma detection system, for example, a performance metric like accuracy treats false negatives and false positives as equally important. This reflects a judgment about the relevant risks of melanoma going undiagnosed versus healthy moles incorrectly being flagged as melanoma.

\begin{figure*}%[ht!]
    \centering
    \includegraphics[width=\textwidth]{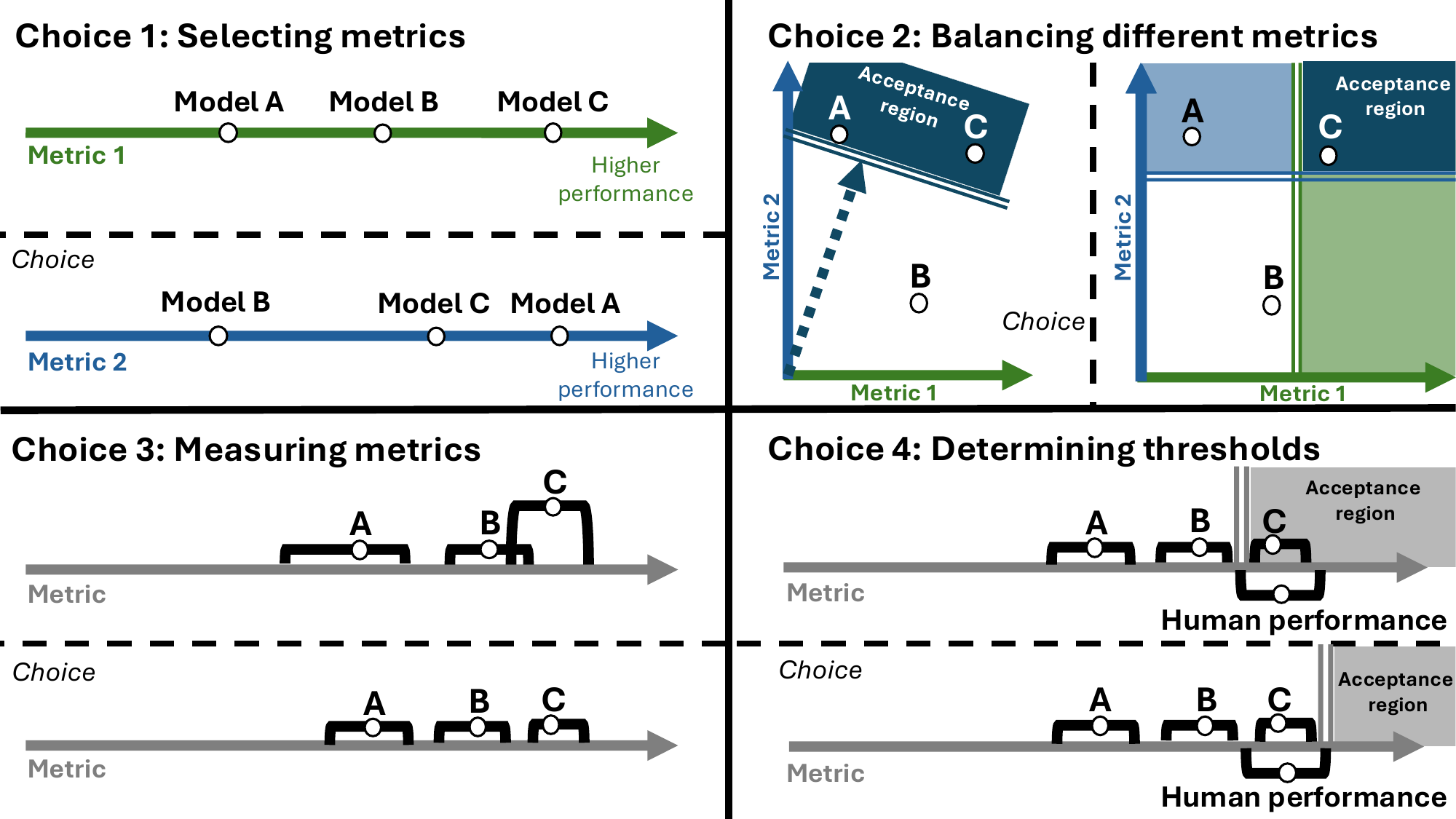}
    \caption{Illustration of the four techno-normative choices of performance evaluations, as examined in this paper. Depending on (1) the chosen metric(s), (2) how they are balanced, (3) how they are measured, and (4) the way in which the acceptance region is defined, different models may be considered more or less accurate. These four choices are thus both central to performance evaluations and directly relevant to the AI Act's requirement for high-risk AI systems to ``achieve an appropriate level of accuracy''. %On the top left is Choice 1: Selecting metrics, explored in Section \ref{sec:metric_choice}; on the top right is Choice 2: Balancing different metrics, explored in Section \ref{sec:balance}; on the bottom left is Choice 3: Measuring metrics explored in Section \ref{sec:measuring}; on the bottom right is Choice 4: Determining thresholds, explored in Section \ref{sec:threshold_choice}
    }
    \Description{Figure with four quadrants each of which represents one of the four techno-normative choices made during evaluation presented in this paper}
    \label{fig:techno_normative_analysis}
\end{figure*}

We refer to such choices as \textit{techno-normative choices}, as both the choice and its technical implementation determine how these normative assumptions materialise in practice. These normative dimensions can easily be taken for granted in data science and ML research \cite{passiProblemFormulationFairness2019, mittelstadtEthicsAlgorithmsMapping2016}.
Insufficient attention to the implications of these choices, however, can have serious consequences for affected persons, particularly in high-stakes domains such as healthcare, public administration, and law enforcement. 
% Our research demonstrates that the specification and operationalization of the problem are always negotiated and elastic, and rarely worked out with explicit normative considerations in mind.

Against this backdrop, the open-ended formulation of the AI Act’s accuracy requirement raises difficult questions. What then counts as an \textit{appropriate} level of accuracy? And how can this be meaningfully assessed across a diverse range of high-risk use cases? At present, neither legal guidelines, nor technical practices provide clear answers to these questions.
%Yet, the open-ended formulation of this requirement, raises difficult questions when viewed in light of the complex technical realities of performance evaluations. What counts as an \textit{appropriate} level of accuracy? And how can this be meaningfully assessed across a diverse range of high-risk use cases? At present, neither legal guidelines, nor technical practices provide clear answers to these questions.  

On the technical side, an extensive body of research has examined the challenges of performance evaluations \cite{derczynskiComplementarityFscoreNLP2016, japkowiczWhyQuestionMachinea, maier-heinMetricsReloadedRecommendations2024, raschkaModelEvaluationModel2020}, 
yet robust performance evaluations in real-world settings remain challenging \cite{hutchinsonEvaluationGapsMachine2022a, passiProblemFormulationFairness2019, lavazzaCommonProblemsUsage2023}.
%[include this..:] % Plus, relatively little technical work has examined how performance evaluations intersect and align with regulatory expectations (cite), with some limited exceptions in medical domains(cite?).
A key source of this difficulty is the mismatch between the dominant learner-centric evaluation paradigm in machine learning (focused on comparing models to one another) and the application-centric perspective often required in high-risk deployment contexts, where performance depends on broader system-level considerations \cite{hutchinsonEvaluationGapsMachine2022a}. The AI Act's accuracy requirement implicitly calls for the latter, while evaluation practices in the ML literature continue to largely reflect the former \cite{hutchinsonEvaluationGapsMachine2022a, derczynskiComplementarityFscoreNLP2016, japkowiczWhyQuestionMachinea}.

On the legal side, there currently are no adopted standards, benchmarks, guidelines, or case law, that help clarify what constitutes an ``appropriate level of accuracy'' under the AI Act. Moreover, the accuracy requirement has so far received little focused attention in legal scholarship. 
Although work is underway on EU harmonised standards that intend to further operationalise the AI Act's requirements \cite{CEN_CENELEC_JTC21_WorkProgramme_2025}, the statistical complexity of AI models and diversity of application contexts make this process challenging. Moreover, standardisation alone is unlikely to resolve the underlying normative and context-specific complexities of performance evaluations \cite{laux2024three}.

Taken together, these gaps create a risk that the AI Act's accuracy obligation will prove difficult to implement, evaluate, and enforce in practice.
As a result, compliance could default to more superficial, learner-centric performance evaluation approaches, where insufficient attention is given to the underlying techno-normative choices involved and the potential harms of a model. %and  potentially resulting in an inadequate assessment of the potential harms that could be caused by deploying the model. %Something like this? 
%with misalignments between technical and legal experts masked by the guise of the objectivity of performance evaluations.  
%Plus, without sufficient attention to the importance of this requirement, there is a risk that compliance may default to more superficial, learner-centric optimisation strategies, which may not sufficiently engage with the (implicit) techno-normative choices embedded in performance evaluations, 

This paper aims to address these challenges by strengthening the interdisciplinary understanding required for an effective implementation of the AI Act’s accuracy requirement. 
%Our aim is to help legal practitioners, auditors and regulators to recognise the legal significance of seemingly technical evaluation choices, and to provide insights on how these choices can/should/may? shape the assessment of wheter an AI system achieves "appropiate level of accuracy", 
% ..consequences that these evaluation choices can now carry? 
To this end, we conduct a systematic analysis of four techno-normative choices that are both central to conducting performance evaluation and most directly impact the accuracy requirement and its associated documentation obligations (as shown in Figure \ref{fig:techno_normative_analysis}): 
\begin{enumerate}[label=\textbf{Choice \arabic*:}, align=left, leftmargin=5em]
    \item Selecting metrics (Section~\ref{sec:metric_choice})
    \item Balancing different metrics (Section~\ref{sec:balance})
    \item Measuring metrics (Section~\ref{sec:measuring})
    \item Determining thresholds (Section~\ref{sec:threshold_choice})
\end{enumerate}
For each choice, we examine its technical role, its relevance under the AI Act, and associated techno-normative complexities. Throughout this analysis, we refer to practical and real-world examples of a high-risk AI use (melanoma detection), and highlight the implications for the implementation and oversight of the AI Act's accuracy requirement. 

%\textbf{Structure.} The remainder of this paper is structured as follows. Section 2 introduces relevant background on the AI Act, and outlines the accuracy requirement in more detail. 
%Section 3 discusses how the accuracy requirement relates to performance evaluations in machine learning practice.
%Following, sections 4–7 each examine one of the four techno-normative choices in turn, and section 8 concludes by discussing the broader implications for AI governance and regulation.
%\textbf{Limitations.} should we not include this in intro, but move this inside the legal section? --> this is sort of done now, but can also keep here in short maybe?  
% Finally, examples used in this paper priamarily focus on supervised learning systems for medical triage in melonoma...
% , chosen for analytical clarity and beacuse of abundant ML research. 
% Yet, we argue that similar choices and complianec challneges arise in other ML tasks, often in even more complex forms..

\section{Accuracy in the EU AI Act}
This section examines the EU AI Act's accuracy requirement in more detail. The accuracy requirement is set out in Article 15(1), and states that:
\begin{quote} 
    ``High-risk AI systems shall be designed and developed in such a way that they achieve an appropriate level of accuracy, robustness, and cybersecurity, and that they perform consistently in those respects throughout their lifecycle.''
\end{quote}
This requirement is however, not a stand-alone obligation. It forms part of a broader, interrelated, framework of requirements for high-risk systems, including substantive documentation duties, and conformity assessment procedures. We therefore first introduce the relevant elements of that framework, before examining the specific function of the accuracy requirement, and ongoing efforts to operationalise it. 

\subsection{The AI Act's High-Risk Requirements, including accuracy}
The EU AI Act, adopted in 2024, aims to ensure that only safe and trustworthy AI systems are used within the EU \cite{EU_2024_AI_Act}. 
The regulation follows a risk-based approach, under which AI systems are classified into different risk categories, with progressively stricter legal obligations. This paper focuses on the \textit{high-risk AI systems} category, as this is the only category subject to a direct legal obligation for accuracy. 

An AI system is classified as high-risk when it can have a significant impact on the health, safety, or fundamental rights of people, which under the AI Act applies in two cases: First, when a system falls within a predefined list of high-risk use cases,  covering systems in education, law enforcement, public services and other high-risk contexts (as included in Annex III). Second, when a system constitutes a product (or a safety component of a product), which is regulated under existing EU product-safety legislation, in the case that it is subject to a third-party conformity assessment prior to being placed on the market. This, for example, applies to MRI-scanners under the EU's Medical Device Regulation (MDR). 
%The AI Act thereby provides a few limited exemptions to the high-risk classifcation, for example when an AI system only performs a narrow procedural, or preparatory task to a final assessment (Article 6(3)). 

The accuracy requirement is part of a comprehensive set of requirements that apply to these high-risk systems. 
Beyond the core performance-related requirements set out in Article 15 (for accuracy, robustness, and cybersecurity) that must be maintained through a system's lifecycle, providers must also implement a risk management system (Article 9), ensure the quality and relevance of training, validation and test data (Article 10(3), mitigate possible bias risks (Article 10(5)), ensure transparency (Article 13), and enable effective human oversight (Article 14(1)).

The responsibility for compliance with these requirements lies primarily with \textit{providers}, which are the entities that develop an AI system and place it on the market.
In most cases, providers may self-assess conformity with the AI Act, provided they create and maintain extensive technical documentation that demonstrates this (Article 11). 
Certain systems, however, are subject to a mandatory third-party conformity assessment, conducted by designated EU notified bodies.
Following market entry, compliance is monitored by national market surveillance authorities, which, starting from August 2027, may impose fines up to 15 million for non-compliance with the high-risk requirements.

When high-risk AI systems are used downstream by deployers (i.e., entities using a system under their own authority), providers must create instructions of use for deployers. Deployers are then responsible for ensuring that the system is used in accordance with those instructions and have to monitor its performance (Article 26).

\subsection{The Function of the Accuracy Obligation}
Although accuracy has a precise technical meaning in machine learning (namely: a performance metric that refers to the proportion of correct predictions over all predictions, see Table~\ref{tab:ai-metrics-legal}), the term serves a different function in the AI Act. 
Rather than referring to a specific statistical metric, accuracy operates as a broader umbrella concept for system performance or functional correctness, and is explicitly linked to a system's intended purpose. 

This functional understanding is made explicit in the European Commission's request to CEN-CENELEC for the creation of EU harmonised standards for the AI Act. This document clarifies that accuracy refers to ``the capability of the AI system to perform the task for which it has been designed'', and should not be confused with `statistical accuracy' \cite{EU_Commission_request_standards}. The standardisation request further emphasises that accuracy assessment must take into account the risks that systems may pose to health, safety or fundamental rights.

The purpose of the accuracy requirement is therefore not to prescribe a specific metric or numerical threshold. Instead, it aims to ensure that providers assess whether their systems perform to an acceptable degree in light of their intended purpose, given the specific deployment context and foreseeable risks. 
As further clarified in Recital 74, this assessment must be made in accordance with the general state of the art and maintained throughout a system’s lifecycle.

This broader, context-specific approach of the accuracy requirement is also reflected in the technical documentation requirements related to accuracy. These state that providers must list the different metrics used to measure accuracy, justify their appropriateness for the specific system, and indicate both the relevant degrees of overall expected accuracy and accuracy levels for relevant persons or groups (Annex IV(2--4)).

\subsection{Ongoing Efforts Towards Operationalising the Accuracy Obligation: Technical Standards and EU Benchmarks}
The AI Act does not specify what constitutes an ``appropriate level of accuracy'', nor how such a level can be determined or justified in practice. 
This design reflects the general approach of EU product safety legislation within the EU New Legislative Framework (NLF), under which legislations are intended to lay down only so-called `essential requirements'. These requirements should specify the results to be achieved, or risks to be addressed, without detailing any technical solutions for doing so \cite{blue_guide}. Technical implementation details are left to secondary mechanisms \cite{blue_guide}, most notably EU harmonised standards. This allows legislation to remain technology-neutral and future-proof, as also intended by the AI Act.

Under the NLF, the wording of essential requirements should nevertheless be precise enough to allow for the development of technical specifications and, in their absence, providers and authorities should still be able to assess compliance \cite{blue_guide}. 
While this approach has proven workable in many traditional product sectors \cite{laux2024three}, it is more complex in the context of the AI Act, and in particular with respect to the open-ended formulation of the accuracy requirement.
Compared to sectors such as machinery, elevators, or medical devices, 
the complex statistical nature of AI models, and the context-specific and normative dimensions embedded in performance evaluations make it considerably more difficult to specify and assess what counts as an ``appropriate'' level of accuracy. 
These difficulties form a central motivation for this paper.

Beyond the EU's product regulations, past literature has also analysed the accuracy principle under the EU's data protection laws \cite{fusterInaccuracyPrivacyenhancingTool2010a}, yet its focus here is different from the performance-related focus, rendering a detailed comparison difficult and beyond the scope of this paper.
%The context-specific and normative dimensions of performance evaluations of AI systems is a central motivation for this paper. 

Against this backdrop, the AI Act relies on two primary mechanisms to support the practical operationalisation of the accuracy requirement: EU harmonised standards and the development of benchmarks and measurement methodologies. We outline the role of each mechanism, and the challenges that both approaches face in light of the accuracy requirement. 

\paragraph{Harmonised standards}
The AI Act primarily relies on harmonised standards to operationalise its essential requirements. These standards are developed by European standardisation organisations -- CEN-CENELEC in the case of the AI Act -- following a formal standardisation request from the European Commission. Once adopted in the EU's Official Journal, organisations can use these standards to demonstrate compliance, which gives them a presumption of conformity with the corresponding legal obligations \cite{ECHarmonisedStandards}. Their use, however, remains voluntary, and organisations may rely on other means to demonstrate compliance \cite{blue_guide}. Similarly, standards do not replace legally binding essential requirements, and organisations remain fully responsible for meeting those \cite{blue_guide}. 

Work is currently underway on a comprehensive set of harmonised standards for the AI Act.
The performance-related requirements for accuracy and robustness under Article 15 are expected to be addressed within a shared standard, called (\textit{the AI trustworthiness framework – Part 2: Accuracy and robustness (prEN 18229-2)}). \cite{CEN_CENELEC_JTC21_WorkProgramme_2025, LeonSmith_programme, draft_trustwothiness_CEN_CENELEC_prEN18229_2_2025}. This framework intends to set out mechanisms for assessing the performance (or: functional correctness) of an AI system, including how to select, measure and validate accuracy metrics, and what to document to meet compliance \cite{EU_Commission_request_standards,draft_trustwothiness_CEN_CENELEC_prEN18229_2_2025}. 

This standard is however unlikely to resolve the core normative and context-specific complexities involved in performance evaluations, as discussed in this paper.
EU standardisation processes typically refrain from addressing hard normative questions in standards \cite{laux2024three}. Plus, the diversity of AI use cases, combined with a general lack of scientific consensus on acceptable thresholds, error rates, or trade-offs across domains, would make such an exercise incredibly complex. These limitations are also reflected in existing international standards on software testing and ML evaluation, which prioritise procedural guidance over detailed advice on acceptable performance levels and context-specific considerations (\cite{isoiec_ts4213_2022, nist_ai_rmf_playbook_measure}).
This problem could potentially be addressed in more task-specific standards, and work is underway for two AI Act relevant task-specific technical reports related to accuracy (in computer vision and natural language tasks) \cite{CEN_CENELEC_JTC21_WorkProgramme_2025}. Still, such standards are equally unlikely to offer explicit guidance on context-specific risks and normative choices in model performance evaluations.

As a result, harmonised standards for the accuracy requirement will likely focus on supporting procedural compliance, while leaving key techno-normative choices to providers, and ultimately, legal assessment by enforcement agencies. 

\paragraph{Benchmarks and measurement methodologies}
In addition to harmonised standards, Article 15(2) introduces a second mechanism,  specific to the accuracy requirement. This states that the European Commission shall encourage the development of benchmarks and measurement methodologies to address the technical aspects of measuring the appropriate levels of accuracy and robustness.

At present, however, it remains uncertain whether such benchmarks are being developed, what form they would take, or how these would be able to address the context-specific and normative dimensions of accuracy evaluations across domains.
% fix.. short
%While benchmarking plays a important role in some other product ares, such as .. , applying this approach to AI systems again likely raises challenges..
%Still have to cite some work here I think? That talks about such benchmarks for other product laws (phsyical prodcut safety things, like fial 1/10000 times for vacuum cleaners), where this might be easier (use eu metrology law (cite arnoud engelfiret)

Taken together, 
%although both mechanisms play an important role in further operationalising the accuracy requirement and related documentation obligations, both mechanisms face substantial difficulties and limitations. 
these difficulties highlight the need for more interdisciplinary work on how accuracy is assessed in practice and whether and how the techno-normative choices embedded in performance evaluations can be meaningfully addressed under the AI Act.

\section{From Legal Requirements to Technical Practice}
\subsection{Accuracy and performance evaluations in ML workflows}
In machine learning practice, questions of whether an AI model achieves an appropriate level of performance are primarily addressed during the performance evaluation stage in the ML development process. 
The goal of performance evaluations is to produce quantitative evidence (typically scores on performance metrics) that are used to assess whether a trained model meets the objectives set out for its task \cite{hutchinsonEvaluationGapsMachine2022a, ISO29119_11_2020}. 
%As illustrated in Figure \ref{fig:workflow}, 
%This stage is commonly situated between model development and deployment, and thus constitutes the final step at which a system's readiness for deployment is assessed \cite{}. 
% \begin{figure}
%     \centering
%     \includegraphics[width=0.75\linewidth]{figures/ML-workflow-highlight.png}
%     \caption{Overview of a standard ML development workflow, highlighting the performance evaluation stage at which quantitative evidence (typically scores on performance metrics) is generated to assess a system’s readiness for deployment. This stage is therefore most directly implicated by the AI Act’s accuracy requirement (adapted from ISO/IEC TR 29119-11:2020 \cite{ISO29119_11_2020}).}
%     \label{fig:placeholder}
% \end{figure}

From a legal perspective, performance evaluation is therefore also the stage that is most directly impacted by the AI Act's accuracy requirement. It is the point at which a system's performance is measured, interpreted, and assessed against standards of acceptability that are linked to a system's intended purpose, 
and where the decision is made whether the system may be deployed. As such, it also plays a crucial role in assessing whether relevant risks have been adequately identified and addressed.

At the same time, performance evaluations are not an isolated step. They are shaped by choices made throughout the entire model development process, including how a system's intended purpose is defined, how data is collected, processed, and split, and how different candidate models are selected and trained. 
Many of these more upstream aspects are, however, more directly addressed by other high-risk requirements in the AI Act, such as those on data governance, risk management, or robustness. 
While the interaction between these requirements and the accuracy obligation is important for compliance in practice, analysing these complex interdependencies is beyond the scope of this paper. 
%Understanding how these overlapping legal and technical frameworks can jointly shape performance assessment and accountability requires a deeper analysis, but in this paper, we will focus on the accuracy requirement of the AIA in isolation.  

This paper, instead, focuses on four specific choices 
-- selecting performance metrics, balancing metrics, measuring performance, and setting performance thresholds --
that are both central to performance evaluations, and most explicitly relevant for demonstrating compliance with the accuracy requirement and related documentation obligations. 
As we demonstrate throughout the next sections, each of these choices encodes implicit or explicit normative assumptions in a model that can impact deployment risks, and thus influence the assessment of appropriate performance under the AI Act. 
And although these choices are not necessarily confined to the evaluation stage and may be influenced by earlier development decisions, their legal and practical significance becomes most visible at the performance evaluation stage. 

\subsection{Structure of the Analysis and Illustrative Use Case}
The following sections analyse these four techno-normative choices in an order that reflects the typical progression of performance evaluations in practice: from the selection of metrics, through the aggregation and measurement of those metrics, to the evaluation of those metrics against thresholds, which ultimately determines the deployability of a system. Each section examines the technical aspects of the choice, its legal relevance under the AI Act, and the normative assumptions embedded.

In each choice section, we draw on concrete examples from a shared high-risk use case: an AI-based melanoma detection system. This use case was chosen because such systems operate in high-risk contexts, where these choices can have serious consequences. Plus, such systems would likely to qualify as high-risk AI systems under the AI Act, as they can be considered as medical devices subject to a third-party conformity assessment under the EU's Medical Device Regulation. 
Moreover, there exists extensive ML research on melanoma detection \cite{dreiseitlComputerHumanDiagnosis2009, phillipsAssessmentAccuracyArtificial2019, phillipsDetectionMalignantMelanoma2019, dickAccuracyComputerAidedDiagnosis2019, ferrisEarlyDetectionMelanoma2021, brinkerComparingArtificialIntelligence2019, haenssleManMachineDiagnostic2018, naeemMalignantMelanomaClassification2020, marsdenAccuracyArtificialIntelligence2024, hosseinzadehkassaniComparativeStudyDeep2019, jojoaacostaMelanomaDiagnosisUsing2021}, alongside increasing real-world adoption, on which we can draw for examples. 

In addition, melanoma detection systems are commonly developed using supervised learning techniques (in which performance is compared against gold-standard human-labelled data), which simplifies our analysis, as this allows performance to be quantified concretely as a function of the similarity between predictions and gold-standard labels.
Similar choices and related challenges, however, arise in other ML tasks, often in more complex forms.

Importantly, our use of real-world examples is not intended as a means to criticise specific systems or practices, but rather to illustrate the complexities and wide range of different approaches that can be taken in performance evaluations, and the impact of these on compliance with the AI Act.
% dont knwo where else to put
In addition, we do not analyse in this paper how sector-specific EU requirements, such as those for medical devices in the case of melanoma detection, can further shape the implementation and assessment of the accuracy requirement.

\section{Choice 1: Selecting Metrics}\label{sec:metric_choice}
\subsection{The choice and its relevance for the AI Act}
The purpose of performance evaluations is to produce evidence of how well an AI model performs relative to its intended purpose. This evidence may take different forms, but in most cases, it is generated through automated, quantitative evaluations, for which performance metrics are used. Such metrics intend to measure how closely a model’s outputs align with desired outputs, based on test data that reflects the system's intended deployment context. 
Selecting one or more performance metrics is therefore a necessary first step in any performance evaluation. 

Metric selection, however, is not a straightforward technical task. There exist a wide range of performance metrics, which are widely discussed in technical literature \cite{hutchinsonEvaluationGapsMachine2022a,maier-heinMetricsReloadedRecommendations2024, ferriExperimentalComparisonPerformance2009}.  
%Even for a single task type, such as classification, a technical standard includes over X different metrics (cite standard). 
Every one of these metrics captures different aspects of model performance and reflects different assumptions. 
For example, in classification tasks, the commonly used \textit{Accuracy} metric assigns equal importance to false negative and false positive errors, whereas the \textit{Precision} metric, in contrast, disregards false negatives altogether, see Table~\ref{tab:ai-metrics-legal}.   
The suitability of any metric, therefore, depends not only on the specific task, but also on your data, and deployment context. 
Table~\ref{tab:ai-metrics-legal} shows some of the most widely used metrics.

\begin{table*}
    \centering
    \footnotesize
    \caption{Definitions and Implications of Commonly Used AI Performance Metrics}
    \label{tab:ai-metrics-legal}
    \renewcommand{\arraystretch}{1.5}
    \resizebox{\textwidth}{!}{%
    \begin{tabularx}{\textwidth}{@{}l >{\RaggedRight}X >{\RaggedRight}X@{}}
        \toprule
        \textbf{Metric} & \textbf{What It Measures (Definition)} & \textbf{What It Means (Implications)} \\
        \midrule
        
        \textbf{Accuracy} & 
        The percentage of correct predictions out of the total number of predictions made. & 
        \textbf{Can be deceptive.} In datasets where one outcome is rare (e.g., melanoma), a model can have high accuracy by simply ignoring the rare cases. \\

        \textbf{Precision} & 
        The quality of a positive prediction. When the AI claims something is true (e.g., ``This is a risk''), how often is it actually true? & 
        \textbf{Minimises False Positives.} High precision ensures the system does not flag innocent data or harmless events. Low precision can lead to user distrust, unnecessary blocks, or excessive use of resources.\\

        \textbf{Recall} & 
        The coverage of actual positives. Out of all the positive cases (e.g., melanoma), what percentage did the AI manage to find? & 
        \textbf{Minimises False Negatives.} High \textit{Recall} is essential for safety-critical AI (like melanoma detection) to ensure no dangers are missed, even if it causes some false alarms (by compromising on precision). \\
        %\midrule

        \begin{tabular}[t]{@{}l@{}}\textbf{F1-Score} \\ \textit{(Aggregated Metric)}\end{tabular} & 
        A combined score that balances \textit{Precision} and \textit{Recall} into a single number. & 
        \textbf{Balances Precision and Recall.} Useful when the AI needs to balance caution (\textit{Precision}) with safety (\textit{Recall}) but hides the specific error type.\\

        \begin{tabular}[t]{@{}l@{}}\textbf{AUROC} \\ \textit{(Aggregated Metric)}\end{tabular} & 
        A measure of the model's ability to separate classes. It asks: ``How likely is a random positive sample to be given a higher classification probability of being positive than a random negative sample '' & 
        \textbf{Goes beyond binary predictions}. Unlike \textit{Accuracy}, it considers the model's continuous scores across different thresholds. However, it can be deceptive in cases of extreme class imbalance.  \\
        \bottomrule
    \end{tabularx}
    }
\end{table*}

Intuitively, choosing a performance metric involves determining which differences between an ideal model, and a trained model are considered most important. Metric selection thus entails weighing the relative consequences of different types of errors, which reflects important normative judgements about the acceptability of errors, harms, or prioritisations. This can have important consequences in the case of high-risk systems. 
%for affected persons, and the overall usefulness and safety of a system. 

For this reason, metric selection also forms a central part of the AI Act’s accuracy requirement. The technical documentation obligations related to the accuracy requirement require providers of high-risk systems to specify which metrics are used to measure accuracy, and to justify their appropriateness (Annex IV(4)). 
Similarly, established international standards for software testing and AI evaluation, such as 
ISO/IEC TR 29119-11:2020 on software testing, explicitly call for justifying the choice of metrics based on risks of a model. 
Performance metrics thus function as one of the most important elements for demonstrating compliance with the accuracy requirement.

Yet, at the same time, neither the AI Act, nor existing technical standards currently prescribe which metric should be used in specific contexts, or what form a justification should take. 
As a result, a meaningful implementation of the accuracy requirement depends on how much attention and scrutiny are given to this choice in practice, and how its normative complexities are addressed. 
To further illustrate the impact of metric selection in practice, we next examine the complexities and normative dimensions of this choice in the context of a high-risk AI-based melanoma detection system. 

\subsection{Technical considerations: Metric selection in practice}
\label{subsec:metric_challenges}
% Introduce chapter and structure
The selection of performance metrics is a highly consequential choice in melanoma detection tasks, as it directly shapes how some important deployment risks are taken into account by a model. The choice is also rarely straightforward, and in practice, many strategies can be observed. 

Two context-specific characteristics are particularly influential for selecting metrics in melanoma detection. 
First, as discussed in the introduction, melanoma detection tasks are characterised by strong class imbalance, as the vast majority of moles in a population are not malignant. Second, the costs associated with different types of errors are asymmetric, as different possible mispredictions entail different forms of harm. Together, these complicate the use of simple aggregate metrics, such as overall accuracy, and require more deliberate evaluation choices. 

Importantly, these characteristics are not unique to melanoma detection, and arise in many other high-risk AI contexts, including fraud detection, recidivism prediction, and AI-based safety components in critical infrastructure, where there are again strong class imbalances (e.g., most people do not commit fraud) and differing harms related to the different types of mispredictions (e.g, safety components that fail to intervene versus false interventions).

\paragraph{Class imbalanced datasets.} 
When metrics assign equal weight to all samples -- such as \textit{Accuracy}, as used in the example of melanoma detection in the introduction -- overall performance scores are largely driven by performance on the majority class, potentially masking poor performance on the rare, yet often more critical, malignant class. 

One approach to solve this, particularly in multi-class settings (for example when trying to identify the type of melanoma), is to compute and report performance metrics separately for each class, allowing errors to to be assessed independently for each class, as can be observed in \cite{hosseinzadehkassaniComparativeStudyDeep2019}. However, while this approach is feasible with certain metrics like \textit{Precision},  \textit{Recall} or \textit{AUROC} (see Table~\ref{tab:ai-metrics-legal}), it does not apply to \textit{Accuracy} since accuracy is equivalent to recall when computed independently for each class. 

An alternative in such cases is to use metrics that normalise the influence of each class on final performance, independently of class size, as is done, for example, with the \textit{Balanced Accuracy} metric~\cite{jojoaacostaMelanomaDiagnosisUsing2021}.
%or  as can be done by applying macro averaging to metrics such as precision, recall or F score. 
Such approaches can help to make visible and mitigate the potential distorting effects of class imbalance. However, they do not eliminate the need for further judgments about the asymmetric cost of misclassification errors, which often plays an important role. 

% Assymetric cost of errors
\paragraph{Asymmetric error costs.} 
%Second, melanoma detection tasks are characterised by asymmetric costs of errors in classification.
In melanoma detection, a system can produce two types of error: (i) false positives, where a benign mole is classified as potentially malignant, which may lead to unnecessary follow-up examinations, and anxiety for patients, and (ii), false negatives, where a malignant mole is classified as benign, which can have more severe consequences, such as a patient not receiving treatment. These two errors are thus not equally harmful, even though they are (implicitly) treated as equally important under commonly used metrics such as accuracy.

% Normativity of choice
Addressing this asymmetry requires determining the relative weight that should be assigned to different kinds of mispredictions. This, in the case of melanoma detection, requires difficult judgements about the severity of false negatives and positives, including how to quantify such differences, which are inherently normative judgements.
Plus, afterwards, a technical challenge remains: how can this normative judgment be encoded into one or more performance metrics? A possible approach is to multiply the penalty given for a type of error (false positive or negative) by a factor $U$ that encodes their relative importance. However, assuming the existence of a valid factor $U$ is in itself a normative assumption that implies the relative importance of an outcome such as a patient's death can be weighed against healthcare system efficiency, a view that can be ethically controversial \cite{herlitzEffectivenessAnalysisValue2025, griffinAreThereIncommensurable1977}

\paragraph{Implications for the AI Act} 
Choosing adequate metrics is thus a crucial step for minimising risks (and thus ensuring compliance with the AI Act's accuracy requirement). As we have shown, making this choice requires challenging normative judgments that are sometimes far from easy to implement technically. This complexity is even more significant when trying to assess the choice from an outside perspective (as potentially done by legal teams, external auditors or enforcement agencies), due to the limited access and resources that may be available. Careful attention is needed to ensure that the normative values behind the choice of accuracy metrics are sufficiently taken into account and documented during the development of high-risk AI systems. 
% Both crucial choices for risks (and thus compliance with appropaite level), hihgly normative, and far from easy technically to implement, let alone to assess from outside perspective (like for legal teams, external auditors, or enforcement). Careful attention needed, and good reflections on judgements, etc.

\section{Choice 2: Balancing multiple metrics} \label{sec:balance}
\subsection{The choice and its relevance to the AI Act}
The evaluation of a model's performance often requires the balancing of different objectives, such as the minimisation of specific types of misclassifications, or balancing the performance on different (sub)tasks performed by the AI model. Each of these objectives may require an independent metric in order to adequately assess a model's performance. When multiple metrics are chosen to evaluate the performance of an AI system, this again leads to an important techno-normative choice: deciding whether and how to aggregate these multiple metrics.

% This can be done in different ways
Aggregating metrics may seem to simplify the accuracy requirement assessment by combining metrics into a single value for which an acceptance threshold can be found. However, the way in which this aggregation is performed dictates (either implicitly or explicitly) how much each metric contributes to the aggregated value. 
Conversely, reporting metrics independently can help ensure that each of the desirable performance metrics is evaluated independently and reaches satisfactory levels. However, this 
introduces complexity by requiring the choice of a threshold for each of the metrics. 

% The choices made in aggregating are normative
In aggregating metrics, we need to determine which aspects of the model's performance we prioritise and to what extent we prioritise them, similarly to how we did in Section \ref{sec:metric_choice}. However, we now make this choice at the metric level rather than at the misprediction level. This remains a techno-normative choice, due to the normativity of prioritising certain aspects of performance over others, and the technical approach through which this choice is resolved.  
While not aggregating metrics may seem to bypass this choice, the choice simply presents itself differently. In setting thresholds, we may be more restrictive for certain metrics than for others, ultimately again prioritising certain aspects of performance over others. 

% So balancing is part of AIA accuracy requirement and related standards
%Balancing metrics affects two aspects of the implementation of the AI Act's accuracy requirement. Aggregation renders the documentation obligations less effective in providing transparency over the normative choices of model evaluation, and the compliance assessment of the requirement becomes more challenging since the normative consequences of the performance are harder to disentangle. The importance of disaggregation is even mentioned in the context of bias evaluation in Annex IV (3). Where the technical documentation requires that \textit{the degrees of accuracy for specific persons or groups of persons} be reported. There is no explicit mention of disaggregation of different performance metrics, but the same underlying reasoning may apply. In that case, the different metrics should be reported independently for increased visibility. 
If a developer were to aggregate multiple metrics together, this could render the documentation obligations less effective in providing transparency over the normative choices made in model evaluations. Similarly, compliance assessment of the accuracy requirement would thereby become more challenging, since the normative consequences of the performance are harder to disentangle.
The importance of disaggregation is already mentioned in the context of bias evaluations in Annex IV (3) of the AI Act. In that case, the technical documentation requires that \textit{the degrees of accuracy for specific persons or groups of persons} be reported. There is, however, no explicit mention of disaggregation of different performance metrics, but the same underlying reasoning may apply. In that case, the different metrics should be reported independently for increased visibility.

% Regulations/standards don't resolve the choice
In practice, many of the most popular metrics (such as \textit{Balanced Accuracy}, \textit{AUROC} or \textit{F scores}, see Table~\ref{tab:ai-metrics-legal}) are aggregations of more granular metrics. Standards acknowledge this reality and even recommend the usage of some of such metrics. But whether aggregation is the right approach to balancing metrics and how this aggregation should be performed remains an open question. 

Through the melanoma detection use case, we next highlight how this techno-normative choice could be made, and its consequences in practice.

\subsection{Technical considerations: Balancing multiple metrics in practice}
The balancing of multiple objectives can be present even in relatively simpler binary classification tasks, such as melanoma detection. Particularly relevant in melanoma detection tasks is balancing a model's \textit{Recall} (what proportion of all melanoma cases get sent to a dermatologist) against \textit{Precision} (what proportion of patients sent to the dermatologist actually had melanoma). Table~\ref{tab:ai-metrics-legal} further describes these metrics. In this subsection, we explore how this balancing can be performed in practice and discuss the implications that this can have for the implementation of the AIA accuracy requirement. 

% To aggregate or not to aggregate
One possible approach to balancing \textit{Precision} and \textit{Recall} is to use an aggregated metric such as \textit{F-score}. \textit{F-score} is a metric that is derived from a non-linear aggregation of \textit{Precision} and \textit{Recall}, known as the harmonic mean. It includes a parameter $\beta$ that allows to control the extent to which \textit{Precision} is prioritized over \textit{Recall} (or the other way around).  However, due to the complex nature of the harmonic mean (especially when $\beta \neq 1$), even if one were to know the degree to which to prioritise one metric over the other, it would not be trivial to practically do so. As the effect of $\beta$ on the relative importance is non-linear. Furthermore, trying to assess the normative values encoded in such a metric post-hoc would require significantly more resources since the specific contribution of each metric is not linear.

Other approaches to aggregation exist that offer simpler ways to balance objectives, for example, using weighted sums of the different metrics. These would allow to more clearly determine what degree of importance was given to the different performance metrics. However, in this case, the complexity increases significantly if many metrics need to be given different weights. 

% Not aggregating
An alternative approach would be to disregard aggregation altogether. This increases the transparency when reporting performance, since it avoids the issue of entangling different metrics in a single value. However, the techno-normative choice of the balancing of metrics does not disappear in this case, it simply reappears in the moment in which we choose acceptance thresholds, as described in Section \ref{sec:threshold_choice}. In the case of melanoma detection tasks, this would mean deciding which thresholds to set for both precision and recall. 
% % How it's done in practice
% Very often papers actually don't decide to aggregate since they are aiming for an exhaustive reporting of the model's performance for compariso sake. 

% Implications for the AIA accuracy requirement
\paragraph{Implications for the AI Act} One of the objectives of the AIA is to ensure adequate documentation of the choices made during the process of developing an AI system to allow for oversight and ensure that risks are taken into account. In addition, providers have to create instructions of use to potential users of their system (i.e., deployers), to make those aware of possible risks and limitations. 
Disaggregating metrics helps with both of these objectives, as it provides more granular information about a model's performance. 
Ideally, such documentation should then not only focus on reporting statistical weights, but also on documenting the normative decisions that were made, and how these are encoded in the model (i.e., how the translation between the two was made). 
% that it not only requires documentation of the statistical weights but also of the human weights (and how the translation between the two was made)

\section{Choice 3: Measuring accuracy metrics} \label{sec:measuring}
\subsection{The choice and its relevance to the AI Act}

% Metrics need to be measured
Beyond metric selection, the measurement process of these metrics also involves a fundamental techno-normative choice. This choice is primarily manifested in two components: \emph{the test set} that is used for measuring, and the \emph{uncertainty estimation approach} that is used for approximating the confidence of this measurement. The objective of this measurement is to provide reliable information on how the model would perform once it is deployed and used on data that it was not trained on.

% There are important choices involved
Selecting the test data on which the model's performance is measured is critical to ensure it can fulfil its intended purpose. The \emph{test set} is a set of data that the model was not trained on which allows to measure the performance on unseen data. The test set must be representative of the real data distribution on which the model will be used, since otherwise, the scores on the performance metrics will not be indicative of the real-world expected performance.

Similarly, the approach for uncertainty estimation should be applied in a way that ensures the estimated confidence interval adequately represents the expected range of performance in plausible real-world deployment distributions. 

% These choices are normative
Both the \emph{test set} selection and \emph{uncertainty estimation approach} thereby depend on the same underlying choice: how to obtain a sample from our dataset that is representative of the real population on which the system will be used. Or, in other words, how to ensure that the variability within and between samples is as representative as possible of real post-deployment subsets? 
To achieve this, the selection of the test set has to ensure that our measure is adequately centred, while the uncertainty estimation approach has to ensure the confidence intervals are actually representative, as shown in Figure \ref{fig:techno_normative_analysis}. 
The normative nature of these choices arises when considering the practical consequences of this choice: if certain subgroups of the population are under-represented in the data samples, both the measurement scores and the related confidence levels may not reliably encompass these subgroups. 

% So the AIA aims to regulate them
The AIA explicitly addresses the importance of the test set choice, both in the accuracy requirement itself and under the data governance requirements in Article 10. The accuracy requirement explicitly addresses the importance of how accuracy is measured in Article 15(2), referring to the work to be done to develop appropriate benchmarks and measurement methodologies.  
The importance of Article 10 comes particularly from 10(3) and 10(4), where the adequacy of the test data for the AI system's intended purpose is discussed. 

% Still choices to be made
While the AIA and its related technical standards discuss the measurement of accuracy more thoroughly than other choices, many aspects still depend heavily on the practitioner's technical implementation. As we will explore in the next section, this technical implementation is not always so simple. 

\subsection{Technical considerations: Measuring metrics in practice}
% Introduction
The issue of adequately measuring metrics is particularly relevant in melanoma detection tasks, since melanoma can manifest itself differently across populations.
In this subsection, we explore the techno-normative complexities involved in selecting a test set and an approach to uncertainty quantification, and conclude by discussing its implications for the AI Act. 

% The test set choice  
In practice, the most common approach for obtaining a test set is \emph{stratified sampling}. This statistical method allows for extracting different data subsets (also referred to as splits) that contain the same proportion of some pre-defined properties in each of them, for example: having the same proportion of male/female, or healthy/unhealthy data points in the training set and test set. Such an approach would complement the requirement to report performance on subgroups in the technical documentation (Annex IV(3)).
However, when using this approach, a complexity arises when there are multiple properties that need to be stratified on, such as in melanoma detection. When a developer, for instance, has to ensure that both gender, ethnicity, age, and medical outcome are stratified, this would require either a very large dataset or an imperfect stratification \cite{sechidisStratificationMultilabelData2011, szymanskiNetworkPerspectiveStratification2017}. Often, imperfect stratification approaches compromise the joint distribution while preserving it for individual properties. While this may seem like a small technical issue, its consequences can be highly impactful, as an imperfect stratification could, for example, lead to imperfect estimations of performance in intersectional subgroups such as ``black, female''. Without visibility into the performance on intersectional groups, discrimination risks may go overlooked \cite{buolamwiniGenderShadesIntersectional2018}. 

In addition to these examples, there may also be less clear-cut properties for which it would be important to ensure equal distributions. Examples of such properties in melanoma detection tasks could be: easy/hard to predict cases, or melanoma images coming from different devices. Failing to stratify on such properties could lead to clinically meaningful failures \cite{oakden-raynerHiddenStratificationCauses2020}

All of these issues can be further complicated when we have multiple images coming from the same patient. In that case, it could be better to split the dataset at the patient rather than image level, since this would give us better construct validity in terms of expected performance on new patients. However, this would make it more difficult to stratify based on properties at the image (rather than patient) level. 

% The normativity of the choice
A developer's approach to stratification can thus be a highly normative choice, as it determines which characteristics of the data should be more adequately represented and prioritised in the test set, which ultimately will also impact the accuracy requirement, as it will affect the test set's composition. 

% Uncertainty quantification
There are two main sources of uncertainty in the model's performance measurement: the confidence of the learned model and the sampled test data on which the performance is measured. There exist approaches for quantifying both these uncertainties, which can provide important insights for deciding on the appropriateness of a model's performance.

Uncertainty arising from the learned model's confidence can be quantified through conformal prediction approaches, which allow us to obtain confidence intervals over predictions \cite{shaferTutorialConformalPrediction2008}; these can then be aggregated across the test set to determine the uncertainty of the overall measured performance.
Uncertainty from the test data can be quantified by using different types of statistical tests. The most common approach in machine learning is \textit{bootstrapping} \cite{kohaviStudyCrossvalidationBootstrap1995}, generating different possible test sets extracted from a given dataset and using the variability in the given measure between samples to estimate the confidence intervals. In practice, this means the model is tested on different realistic possible test sets, which gives a better understanding of the performance range of the model. % There was a comment from the reviewer to add clopper-pearson but not sure whether/how to do it, maybe interesting for practicioners but no need to go in-depth into all possible ways to do uncertainty quantification?

\paragraph{Implications for the AI Act} Through the data and data governance requirements of the AI Act (and the related work on technical standards for data management), the AI Act addresses several important aspects that relate to the measurement of performance metrics. 

At the same time, as we showed through a discussion of measuring performance in melanoma detection tasks, important techno-normative complexities remain when performing such measurements in practice, which can be highly impactful for determining the appropriateness of a system's performance.

To ensure a meaningful implementation and adequate assessments of the accuracy requirement,  attention should thus also be paid to documenting and disclosing the approaches used for selecting test set data, stratification, and uncertainty estimations in the context of a provider's accuracy assessments. 
% \subsection{Implications for the AI Act implementation}

\section{Choice 4: Determining acceptance thresholds for accuracy}\label{sec:threshold_choice}

\subsection{The choice and its relevance to the AI Act}
% We need to determine what the acceptance threshold is, and the previous choices influence this one
After evaluating a model's performance, it needs to be determined whether its performance is sufficient for deployment. But how can an appropriate level of accuracy be determined for an AI system?

% This can be done in different ways, human level performance? the best human or the average human? 
In practice, different approaches can be taken for making this choice. A seemingly simple option would be to compare the performance of the model to that of a human performing the same task. However, not all humans perform equally at a task, so should we consider the performance of an average human or the best-performing human?
Furthermore, while in some scenarios it may be enough for a model to perform almost as well as a human, in others, its performance will need to be significantly higher. This would be further complicated when multiple (disaggregated) metrics are needed to assess performance. In that case, would it be sufficient for the model to outperform the human in one metric? Or all of them? And considering that the measurement comes with some uncertainty, as discussed in Section \ref{sec:measuring}, how should this uncertainty be incorporated into determining an acceptance threshold? 

% This choice is clearly normative
The choice of an acceptance threshold is the most evidently techno-normative aspect of performance evaluation. Going back to the definition of evaluation metrics as measures of the gap between an ideal model and a current model, the appropriate level -- once the metric(s) and how they are measured has been defined -- establishes the acceptable size of this gap. In other words, what degree of harm is tolerable when deploying this model? Is it enough for it to perform as well as an average human? Or should it be better? By how much? 

% Extra paragraph about linking to previous choices? 
The way in which we chose the accuracy metric(s), how to balance them, and how to measure them plays a critical role in this choice. It is the measurement stemming from these choices that will have to be compared against the acceptance threshold. Therefore, the choice of an acceptance threshold must be made taking into consideration the influence of these previous choices: what is the range of the chosen metrics? Do we have multiple metrics and therefore need multiple thresholds? Should we be more restrictive about the performance on certain metrics than others by setting higher thresholds? Should the performance be above the acceptance threshold for all relevant subgroups of the population? What part of the range of possible values of the measurement of the metrics should fall above the acceptance threshold?   

% But if anything, the standards and regulations give the least guidance on this one, so lots of room for interpretation

The choice of an acceptance threshold is at the core of the AI Act's accuracy requirement. 
This can be seen by the reference to \emph{appropriate} levels, which is clearly linked to choosing acceptance thresholds. 
Yet, the open-ended framing of this requirement leaves open the question of how to make this choice in practice. While this ensures that organisations have the flexibility to assess appropriate acceptance levels in light of their specific use case, it also leaves the door open for misguided techno-normative choices that may lead to harm. We next examine how such choices can arise in the context of the melanoma detection use case. 
%but the question of how to make this choice in practice is left to practioners implementing the requirements. nor is this expected be resolved through the development technical standards. and its related technical standards. These recognise the importance of the contextuality of accuracy and, therefore, of acceptance thresholds. 
%While this ensures that practitioners have more flexibility to assess what an appropriate acceptance level of accuracy may be for their specific AI system, it also leaves the door open for misguided techno-normative choices that may lead to harm.

\subsection{Technical considerations: Determining acceptance thresholds in practice}
% Introduction
In assessing the appropriate level of performance for a given AI system, it is first of all important to consider the environment and context in which the system will operate. 
Depending on that context, a natural first step is to then compare the AI system with existing approaches, but these may not always be present or comparable. 
In this subsection, we start by looking at the influence of the deployment environment and the options for comparing performances, we then look at the role of a golden standard for performance, and conclude by discussing the implications for the AI Act accuracy requirement.

In the case of a melanoma detection system, one scenario could be a hospital that has a system by which high-risk patients have yearly check-ups for skin lesions performed by a dermatologist, while the general population simply gets referred when their general practitioner identifies a potentially malignant skin lesion. 
The assessment of the appropriate level in this situation depends on what role the AI system will play. On the one hand, if the AI system is used to substitute the dermatologist check-ups, the requirement would have to be for its performance to be higher than that of dermatologists. On the other hand, if the AI system is used only to complement general practitioner visits for the general population, a performance that is less high may be sufficient.

% Existence or absence of golden standard
When an AI system is intended to replace an existing procedure, an easy option is to compare their respective performances. However, in certain situations, the appropriate level of performance for the AI system may be below that of the existing procedure. This may be the case when the AI system is used, for example, as a triage mechanism to more adequately allocate the resources required to the patients who may have skin lesions that are more likely to be malignant. 
If we consider some of the criticisms of AI systems, such as the (in)contestability of their decision-making, the resources required to train them, or their potential biases, one might even argue that their performance should be higher in certain cases \cite{mittelstadtEthicsAlgorithmsMapping2016}. 

% When golden standard is absent
If an AI system is used for a new procedure altogether, the selection of an appropriateness level becomes more challenging. In that case, we need to analyse what the potential harms are and whether they are acceptable. This requires a higher degree of normative decision-making. 

\paragraph{Implications for the AI Act} Risks and harms are at the core of the AI Act, and for the accuracy requirement, their magnitude can arguably be most strongly controlled through the acceptance threshold. This makes this choice seemingly the most important to analyse when assessing compliance with the accuracy requirement. However, as we have shown, many aspects can influence the making of this choice, and these aspects are often highly dependent on the task and deployment context. It is therefore crucial that providers adequately document the considerations made in making this choice. Most importantly, the determination and setting of an appropriate threshold can only be properly understood and evaluated by taking into account the three techno-normative choices presented above. 
% The decision of a threshold encodes the tolerable degree of risks and hamrs from an AI model. 

\section{Conclusions: How to Reach More Meaningful Performance Evaluations} \label{sec:discussion_conclusion}
When is an AI model sufficiently accurate to be deployed in high-risk contexts? As we have shown, this question cannot be answered by reference to a single metric or numerical threshold. Under the EU AI Act, accuracy also does not function as a narrow technical property, but instead as a context-dependent requirement for system performance, to be assessed in light of a system’s intended purpose and relevant deployment risks. 
 
In practice, the assessment of whether a system's performance is accurate enough is shaped most strongly by a set of techno-normative choices made during ML performance evaluations. These include decisions about metric selection, the balancing of multiple metrics, measurement procedures, and the setting of acceptance thresholds. As we illustrated through practical examples from a melanoma detection use case, each of these choices encodes assumptions about acceptable errors, risks, and harms into a model. Therefore, these choices are highly consequential, and require sufficient regulatory oversight, both in the AI Act, and in emerging AI regulations that intend to regulate for accuracy. 
% As we also showed, making such choices is not a straightforward task. It requires engaging with both technical constraints and normative judgments. In practice, this is not always done sufficiently, and, some decision, while seemingly technical at first glance, can, without sufficient reflection, lead to important risks when the model is deployed.

% In this paper, we unveiled the complexity of four foundational techno-normative choices made during performance evaluations. By analyzing these choices both through their interactions with the AI Act and their technical implementation, we bridge some of the gaps in understanding between regulation and technical practice. 
% Crucially, our analysis highlights the need to openly and thoroughly discuss essential techno-normative choices involved in model evaluation. These discussions: require more technical talent and research \cite{reuelOpenProblemsTechnical2025}, must be contextual \cite{detroyaMisabstractionSociotechnicalSystems2025} and should embrace disagreement \cite{fazelpourValueDisagreementAI2025, daiEmbracingContradictionTheoretical2025}. 
% A techno-normative approach to performance evaluation may require significant resources both for AI development and compliance assessment. But if we want high-risk AI systems that are accurate in a way that matters, it is an investment worth making.  

As we showed, making such choices is not an easy task, especially in high-risk contexts. It requires engaging with both complex technical constraints and difficult normative judgments, which are highly context-dependent. 
These complexities also help explain why the AI Act’s accuracy requirement is necessarily open-ended. Given the diversity of high-risk use cases in the AI Act, and the normative complexity inherent in performance evaluation, prescriptive metrics or uniform thresholds cannot work. 

The central challenge for the AI Act therefore lies in implementation: ensuring that the techno-normative choices embedded in performance evaluations are made intentionally, with awareness of the deployment context \cite{detroyaMisabstractionSociotechnicalSystems2025},
with due deliberation, attention to interdisciplinary perspectives, 
and by embracing disagreement and discussion   
\cite{fazelpourValueDisagreementAI2025, daiEmbracingContradictionTheoretical2025}. Similarly, meaningful assessment by external auditors and regulators requires engaging with these underlying techno-normative evaluation choices. Developing the institutional capacity and interdisciplinary expertise needed to support such assessments will similarly be critical \cite{reuelPositionTechnicalResearch2024}. 
%These findings are equally relevant beyond the AI Act, especially for other jurisdictions that seek to regulate the accuracy or performance of AI systems.
%AI regulations beyond the AI Act that aim to introduce accuracy requirements. 

\section*{Generative AI Disclosure}
Google Gemini 3 (Fast, Thinking and Pro) and ChatGPT (GPT-5 and o3) were used for text and grammar editing. Google Scholar Labs was used to find relevant literature.

\begin{acks}
The authors who are affiliated with the "Law \& Tech Lab" are supported by the RegTech4AI AiNed Fellowship Grant, which is funded by the Dutch National Growth Fund (NGF) under file number NGF.1607.22.028. Kristof Meding was supported by the Carl Zeiss Foundation through the CZS Institute for AI and Law.

We extend our gratitude to our colleagues at the Maastricht Law \& Tech Lab for their valuable feedback on earlier drafts of this paper. We also want to thank the organisers and participants of the 2025 Writing Workshop on the European AI Act, hosted by the Chair of Law and AI at the University of Tübingen, for their insightful feedback.
Finally, we are grateful to the anonymous reviewers of the ACM FaccT conference and Mara Seyfert for their thorough feedback on the paper.
\end{acks}

\bibliographystyle{ACM-Reference-Format}
\bibliography{accuracy_bram, accuracy_lucas}

\end{document}